# Phase diagram of the antiferromagnetic *XY* model in two dimensions in a magnetic field

A. S. T. Pires[*] and B. V. Costa[†]

*Departamento de Física, Laboratório de Simulação, ICEX, UFMG, 30123-970, Belo Horizonte, MG, Brazil*

R. A. Dias[‡]

*Departamento de Física, ICE, UFJF, 36036-0900, Juiz de Fora, MG, Brazil*

The phase diagram of the quasi-two-dimensional easy-plane antiferromagnetic model, with a magnetic field applied in the easy plane, is studied using the self-consistent harmonic approximation. We found a linear dependence of the transition temperature as a function of the field for large values of the field. Our results are in agreement with experimental data for the spin-1 honeycomb compound BaNi$_2$V$_2$O$_3$.

The *XY* model in two dimensions provides the best example of a phase transition mediated by topological defects. A variety of analytical and numerical methods have been presented in the literature in an attempt to fully understand the nature of its transition. As it is well known the model has a phase transition at a temperature $T_{BKT}$ called the Berezinskii-Kosterlitz-Thouless (BKT) temperature.[1,2] This phase transition is associated with the emergence of a topological order, resulting from the pairing of vortices with opposite circulation. The BKT mechanism does not involve any spontaneous symmetry breaking[3] and emergence of a spatially uniform order parameter. The low-temperature phase is associated with a quasi-long-range order for finite $T$, with the correlation of the order parameter decaying algebraically in space. Above the critical temperature the correlations decay exponentially. This picture is applicable to a wide variety of two-dimensional phenomena.[4] A recent experiment in a trapped atomic gas[5] not only confirms the BKT theory in a new system, but also reveals for the first time the role played by local topological defects or vortices. A very convenient technique to study the *XY* model is the self-consistent harmonic approximation (SHCA). The SHCA was originally proposed by Pokrovsky and Uimin[6] to study the two-dimensional (2D) classical planar rotor model. Later, Minnhagen[7] pointed out that the SCHA overestimated the transition temperature because it did not take into account vortex fluctuations and he suggested a way to improve the thermodynamics of the planar rotor by replacing the exchange constant $J$ with a "renormalized" $J(T)$. This procedure leads to a better estimate of $T_{BKT}$. Menezes *et al.*[8] extended the SCHA to the classical *XY* model and Pires[9] applied it to the quantum model. The approximation consists in replacing the Hamiltonian of the system by an effective harmonic Hamiltonian with renormalized parameters. Several applications to classical systems were found to agree very well with Monte Carlo and experimental results.[10,11] The SCHA was also used in the study of the 1D quantum sine-Gordon problem, where it describes correctly the phase transition of the model. The reason is that it is equivalent to a renormalization-group analysis to one loop.[12] To test the reliability of the quantum SCHA we will compare the transition temperature calculated theoretically with experimental results, for two $S=1$ compounds.[13] In these cases we have an exchange anisotropy of the form $A(S_i^z S_j^z)$. The calculation proceeds in the same form as presented below and the details are presented in Ref. 14. The first system is the stage-two NiCl$_2$ graphite interlayer compound with $J=20$ K and $\lambda=0.992$. The experimental value for $T_{BKT}$ lies in the region $0.45J$–$0.50J$. The SCHA value is $0.46J$. The second compound is BaNi$_3$(PO$_4$)$_2$ with $|J|=22.0$ K and $\lambda=0.95$. The experimental value is $T_{BKT}=0.523J$ and the theoretical one is $0.505J$. As we see the agreement between the SCHA and experimental data is reasonably good. The SCHA was used to treat the interplane coupling in the planar rotor, where the transition temperature was calculated as a function of the interlayer exchange.[15] The *XY* model with in-plane anisotropy was studied by Spirin and Fridman.[16] The effect of a magnetic field applied perpendicular to the *xy* plane was taken into account by Pires.[14] Recently Knafo *et al.*[17] pointed out that it would be useful if the theoretical models for *XY*-like models, which already consider interplane coupling, would also incorporate in-plane anisotropy. This is the aim of the present Brief Report. The quasi-two-dimensional antiferromagnetic spin-1 compound BaNi$_2$V$_2$O$_3$ is considered to be the best prototype of a quasi-two-dimensional *XY* model.[18] The reported in-plane exchange parameter is $J/k=48$ K. As a consequence of the out-of-plane exchange $J'$, this compound has a phase transition at $T_N=47.4$ K, with the spins aligned in the *xy* plane. Even a small value of $J'$ changes the character of a BKT transition to an order-disorder transition. As far as we know there is no estimate to $J'$. The Hamiltonian we will treat here is written as

$$H = J\sum_{\langle i,j\rangle} \vec{S}_i \cdot \vec{S}_j - J'\sum_{\langle i,j'\rangle} \vec{S}_i \cdot \vec{S}'_j + D_{xy}\sum_i (S_i^z)^2 - \mu_0\sum_i \vec{H}\cdot\vec{S}_i, \quad (1)$$

where $\langle i,j\rangle$ means in-plane neighbors and $\langle i,j'\rangle$ is for out-of-plane neighbors. The single-ion anisotropy, $D_{XY}=0.1J$, although small is quite effective to lead to a *XY* behavior. There is also an in-plane anisotropy $D_{IP}$ quite small, estimated by $4\times 10^{-4}J$, and it will be neglected. $h=H/J$ is a magnetic field applied within the *xy* planes. For $h=0$, the in-plane anisotropy is important for establishing three-dimensional (3D) long-range order. In the honeycomb lattice, this anisotropy acts to align the spins along one of the three equivalent hexagonal easy axes.[19] With no loss of generality we have used a cubic lattice in this work, with a square lattice in the *xy* plane instead of the honeycomb lattice. We have made some preliminary calculations using the honeycomb lattice obtaining qualitatively the same results. However, due to the conditions at the edge of the Brillouin zone, the numerical determination of the solutions of the Eqs. (11)–(13) using the honeycomb lattice demands much more time and much more care, for this reason we decided to use a square lattice in the *xy* plane. The main properties of the model are preserved.

Writing the spins components in the Hamiltonian (1) in terms of the Villain representation,[20]

$$S_n^+ = e^{i\Psi_n}\sqrt{\left(S+\frac{1}{2}\right)^2 - \left(S_n^z + \frac{1}{2}\right)^2},$$

$$S_n^- = \sqrt{\left(S+\frac{1}{2}\right)^2 - \left(S_n^z + \frac{1}{2}\right)^2}\, e^{i\Psi_n}, \quad (2)$$

we obtain

$$H = \frac{J}{2}\sum_{r,a}\left\{S^2\left[1-\left(\frac{S_r^z}{S}\right)^2\right]\cos(\Psi_r - \Psi_{r+a}) + S_r^z S_{r+a}^z\right\}$$

$$-\frac{J_z}{2}\sum_{r,a}\left\{S^2\left[1-\left(\frac{S_r^z}{S}\right)^2\right]\cos(\Psi_r - \Psi_{r+a}) + S_r^z S_{r+a}^z\right\}$$

$$+ D\sum_r (S_r^z)^2 - Sh\sum_r\left[1-\left(\frac{S_r^z}{S}\right)^2\right]\sin\Psi_r. \quad (3)$$

Here *a* stands for the nearest neighbor in a cubic lattice. The external magnetic field is assumed to be aligned along the *y* direction. Following Dotsenko and Uimin[21] we write

$$\Psi_r = \phi_r + \alpha \quad \text{and} \quad \Psi_{r+a} = \phi_{r+a} - \alpha + \pi, \quad (4)$$

where the angle $\alpha$, chosen as to minimize the classical ground-state energy, is given by $\alpha = \arcsin\frac{h}{h_c}$ with $h_c = 8J$. Increasing the field beyond the saturation field $h_c$, all spins align in the field direction. In the SCHA one replaces the original Hamiltonian by an effective harmonic one with renormalized temperature dependent parameters. There is an extensive literature describing the SCHA (Refs. 8, 9, and 14) and, for this reason, we will only sketch the main steps. Following the procedure described, for example, in Ref. 14 one arrives at the following effective Hamiltonian written in terms of Fourier components:

$$H = \sum_q [S^2 a(q)\phi_q\phi_{-q} + b(q)S_q^z S_{-q}^z], \quad (5)$$

where

$$a(q) = \rho_{xy}(1-\gamma_q^{xy})\cos 2\alpha + J_z\rho_z(1-\gamma_q^z) + \frac{h^2\gamma}{16J^2 S}, \quad (6)$$

$$b(q) = D + (\cos 2\alpha + \gamma_q^{xy}) + J_z(1-\gamma_q^z) + \frac{h^2}{16J^2 S}, \quad (7)$$

$$\gamma_q^{xy} = \frac{1}{2}(\cos q_x), \quad \gamma_q^z = \cos q_z, \quad (8)$$

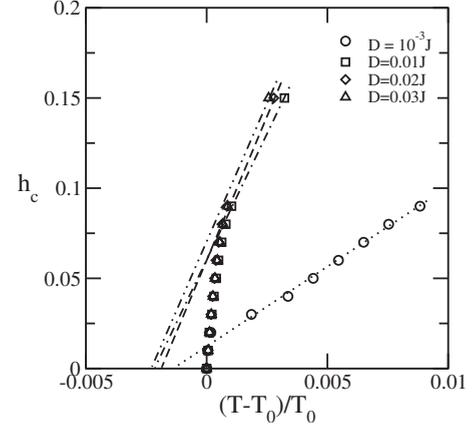

FIG. 1. Phase diagram obtained with the field *h* in the easy plane and several values of the anisotropy *D* (shown in the inset) with $J' = 0.1J$. The low- and high-temperature phases are antiferromagnetic and paramagnetic phases, respectively. The symbols are for self-consistent calculations. The lines are guides for the eyes. Here, $T_0$ is the critical temperature for $h = 0$.

$$\left\langle\left(\frac{S_r^z}{S}\right)^2\right\rangle = \frac{T}{4\pi^3 S^2}\int_0^\pi\int_0^\pi\int_0^\pi \frac{d\vec{q}}{b(q)}, \quad (9)$$

$$R_1 = 1 - \left\langle\left(\frac{S_r^z}{S}\right)^2\right\rangle, \quad R_2 = 1 - \frac{1}{2}\left\langle\left(\frac{S_r^z}{S}\right)^2\right\rangle, \quad (10)$$

$$\rho_{xy} = R_1 \exp -\frac{T}{4\pi^3 S^2}\int_0^\pi\int_0^\pi\int_0^\pi \frac{d\vec{q}(1-\gamma_q^{xy})}{a(q)}, \quad (11)$$

$$\rho_z = R_1 \exp -\frac{T}{4\pi^3 S^2}\int_0^\pi\int_0^\pi\int_0^\pi \frac{d\vec{q}(1-\gamma_q^z)}{a(q)}, \quad (12)$$

$$\gamma = R_2 \exp -\frac{T}{8\pi^3 S^2}\int_0^\pi\int_0^\pi\int_0^\pi \frac{d\vec{q}}{a(q)}. \quad (13)$$

For $h \ll h_c$, $\cos 2\alpha \approx 1$, the spins are aligned nearly perpendicular to the field, and the magnetic field acts as an effective anisotropy. We have the equivalence:

$$D(h) = \frac{g^2\mu_B^2}{16J}h^2. \quad (14)$$

For our case Eq. (14) can be written as $D(h) = 2.34 \times 10^{-3} h^2$(K) with *h* in tesla.

At $J' = 0$, the $T_{BKT}$ temperature is identified as the temperature where the curve intersects the line $2T/\pi$.[7,14] For $J' \neq 0$ the transition temperature $T_N$ is identified as the temperature where $\rho(T)$ drops to zero. There is theoretical and experimental evidence that even for moderate interlayer coupling the universal jump at $T_{BKT}$ is replaced by a rapid downturn of $\rho(T)$ at a temperature above $T_{BKT}$.[22] Figure 1 shows the phase diagram obtained from our calculations. It should be compared with Fig. 3 of Ref. 17. For a question of clarity the horizontal axis was renormalized using the critical temperature, $T_0$, at $h = 0$ for each value of the anisotropy *D*.

Here, $T_0=0.9767$, 1.1554, 1,2197 and 1.2629 for $D=10^{-3}$, 0.1, 0.2, and 0.3, respectively. As we can see $T_N(h)$ is almost constant for $h \lesssim h^*$ and increases linearly for $h \gtrsim h^*$, in agreement with experimental data presented in Ref. 17. Our theoretical calculation describes correctly the experimental data without taking into account the effects of the in-plane Ising-type anisotropy. We remark that Knafo et al.[17] found that $T_N(h)$ was almost constant for $\mu_0 h \leq 2T$ and explained this behavior as due to the effect of the anisotropy DIP and the presence, at zero field, of three kinds of domain in which the spins are oriented along equivalent hexagonal direction. These authors prescribe the linear increase in $T_N(H)$ to a reduction in the spin fluctuations along the direction of the magnetic field, due to the field-induced anisotropy. The transition temperature $T_N(0)$, as calculated here, is the ordering temperature of the system in the case of no in-plane anisotropy. An Ising-type in-plane anisotropy, $D_{IP}$, shifts $T_N$ upward, but since $D_{IP}=4\times 10^{-4}J$, this shift is small. $D_{IP}$ corresponds to an effective magnetic field of $h=0.41T$, smaller than $h^* \approx 2T$.

Layered magnetic compounds exhibit a phase transition to three-dimensional long-range order at a temperature $T_N$, often too large to be exclusively caused by the weak interlayer coupling. It seems that anisotropy is more important to explain the mechanism of ordering in these compounds.[23]

This work was partially supported by CNPq, FAPEMIG (Brazilian Agencies).

*antpires@fisica.ufmg.br
†bvc@fisica.ufmg.br
‡radias@fisica.ufmg.br